\documentclass{aa}  
\usepackage{graphicx}
\usepackage{ulem}
\usepackage{txfonts}
\usepackage{float}
\usepackage{makecell}
\usepackage{threeparttable}
\usepackage{relsize}
\usepackage{longtable}
\usepackage{subcaption}
\usepackage{supertabular}
\usepackage{geometry}
\usepackage{xcolor}
\usepackage{hyperref}
\usepackage{url}
\usepackage{float}
\usepackage{booktabs} 
\usepackage{array}    
\usepackage{multirow}
\usepackage{amsmath}
\usepackage{comment}
\usepackage{textcomp}

\begin{document}

   \title{Identifying lopsidedness in spiral galaxies using a Deep Convolutional Neural Network}

   \author{Biju Saha
          \inst{1}\thanks{e-mail: bijusaha930@gmail.com},
          Suman Sarkar
          \inst{2},
          \and
          Arunima Banerjee
          \inst{1}
          }
    \titlerunning{Identifying lopsidedness in spiral galaxies using DCNN}
    \authorrunning{Saha et al.}
   \institute{Indian Institute of Science, Education and Research, Tirupati 517619, India
         \and
             Indian Institute of Technology, Kharagpur 721302, India\\
             }

    \date{\today}
 
  \abstract
   {About 30\% of disk galaxies show lopsidedness in their stellar disk. Although such a large-scale asymmetry in the disk can be primarily looked upon as a long-lived mode ($m=1$), the physical origin of the lopsidedness in the disk continues to be a puzzle.}
   {In this work, we employ a transfer-learning approach for the automated identification of lopsided galaxies using SDSS DR18 imaging by fine-tuning a Zoobot model, a deep convolutional neural network package pre-trained on the Galaxy Zoo dataset.}
   {We obtain 7,042 well-resolved, nearly face-on spiral galaxies from SDSS DR18 over the redshift range 0.01 $\leq z \leq 0.1$, with extinction-corrected g-band model magnitude < 16 and Petrosian radius (enclosing 90 \% of the flux) $\geq$ 3 arcsec. Out of these, we visually identify 490 lopsided and 444 symmetric galaxy samples suitable for training. The trained model achieves a testing accuracy of $(87 \pm 0.02)$ \%, averaged over 10 independent trials.}
   {Using the best-performing model, we identify 3,679 lopsided and 2,429 symmetric galaxies from the remaining sample. Of these, 2,658 lopsided and 1,455 symmetric galaxies are predicted with are predicted with high prediction probability $P_{pred} \geq 0.85$. Lopsided galaxies in our predicted samples are relatively high star-forming, bluer, low-concentration (late- type), low-mass galaxies compared to the symmetric galaxies.}
  {Our study produces an usable catalogue of lopsided and symmetric galaxies, which will offer new insights into the formation of lopsidedness in disk galaxies. The dataset and the best-performing model are made publicly available through GitHub at \url{https://github.com/bijusaha-astro/CNN_lopsided}}


\maketitle
\section{Introduction}

A significant fraction of disk galaxies host large-scale asymmetry in their gas and/or stellar disk. \citet{Baldwin..1980} was the first to study the asymmetry in atomic hydrogen (HI) distribution of about 20 spiral galaxies where one-half of the distribution appears to be more extended than the other; he coined the term ``lopsided" to indicate such asymmetry in the disk. \citet{Rix..Zaritsky..1995} considered the NIR observations of 18 face-on spiral galaxies and found that 30\% of them are significantly lopsided. They quantified lopsidedness with the $m=1$ amplitude in the Fourier decomposition of the surface brightness distribution. The high incidence of lopsidedness in disk galaxies was further confirmed by (i) \citet{Zaritsky..1997} who studied 60 field spiral galaxies, (ii) \citet{Bournaud..2005} who studied 149 spiral galaxies in the OSUBGS sample, (iii) \citet{Reichard..2009} who investigated lopsidedness in 25,000 nearby galaxies from the SDSS Data Release 4, and (iv) \citet{Zaritsky..2013} who studied a sample of 167 galaxies that span a wide range of morphologies and luminosities, among others. \citet{WHISP..2011} studied the kinematical lopsidedness in a sample of 70 galaxies from the Westerbork H\textsmaller{I} Survey of Spiral and Irregular Galaxies (WHISP). The ubiquity of these large-scale asymmetries suggests that disk is dynamically unstable against $m=1$ perturbation. \\

Lopsidedness influences the dynamics and secular evolution of the galaxy (see \citealp{Jog..2009} for a detailed review). The non-axisymmetric $m=1$ mode exerts gravitational torques on the stars, driving the outward transport of angular momentum (\citealp{Saha..2014}). The possible origin of lopsidedness is either attributed to minor mergers (\citealp{Zaritsky..1997}), tidal interactions (\citealp{kornreich..2002}), or flybys (\citealp{mapelli..2008}). These mechanisms are frequent in denser environments and generate a short-lived (i.e. $\sim$ 1–2 Gyr) but strong lopsided feature in galaxies (\citealp{mapelli..2008}). Interestingly, however, there exists no correlation between lopsidedness and strength of the tidal interaction parameter (\citealp{Bournaud..2005}), or with the presence of nearby companions (\citealp{Wilcots..2004}). Recently, \citet{Dolfi..2023} considered lopsided galaxies sample at $z$ = 0 from the IllustrisTNG simulation and showed that regardless of the environment, the internal properties like the central stellar mass density and disk-to-total mass ratio play a major role in determining the susceptibility of disk galaxies to such perturbations. The above results were further confirmed for redshifts between 0 < $z$ < 2 by \citet{Dolfi..2024}.  However, they observed that at higher-redshift ( 1.5 < $z$ < 2), environmental factors show a significant influence on the mechanisms generating lopsidedness. Strong lopsided features are also observed in late-type field galaxies. \citet{Bournaud..2005} showed that asymmetric gas accretion via cosmological filaments, leading to asymmetric star formation in the disk, can generate lopsidedness even in isolated galaxies. \citet{Saha..2007} demonstrated that a pure exponential stellar disk sustains a global $m=1$ mode and the inclusion of self-gravity makes it long-lived. Through analytical studies, it was demonstrated that a distorted dark matter halo potential (\citealp{Jog..1997,Jog..2000}) or a small offset of 350 pc between the centre of dark matter halo and centre of the galactic disk can lead to strong, long-lived lopsidedness in the disk (\citealp{Prasad..2017}). \\

Despite all these studies, the key generating mechanism and the survival of lopsidedness in disk galaxies is not well understood, especially the occurrence of lopsidedness in isolated galaxies. Though few analytical studies discussed above have been useful in predicting the possible generating mechanism, they rely on linear approximations and require further comparison with simulations. \citet{VL..2023} extracted simulated disk galaxies with Milky Way-like mass halos from the IllustrisTNG simulation and showed that lopsidedness correlates more strongly with internal parameters like central stellar surface density than any particular external driving source. Their results also suggest that lopsided galaxies tend to reside in high-spin and often highly distorted DM halos. Rigorous systematic observational studies coupled with controlled simulation is required to understand the impact of environment, halo and intrinsic properties of the disk in generating lopsidedness. Thus, obtaining a large sample of lopsided galaxies identified from the present and future surveys will contribute towards understanding their origin and evolution.\\ 

With the advent of large volumes of data from the multi-wavelength deep field high-resolution surveys, visually classifying the morphologies of galaxies through the manual inspection of images becomes a laborious task. Further, the high-resolution survey revealed many new interesting features in the morphology, which demanded a more refined morphological classification system. The citizen science Galaxy Zoo 2 (GZ2) project introduced a visual morphological classification for more than 300 000, wherein volunteers classify galaxies by answering a series of questions based on galaxy images (\citealp{Willett..2013}). GZ2 project resulted in the detailed morphological classification of the largest and brightest SDSS galaxies. However, with the huge volume of images from future surveys such as the Rubin Observatory Legacy Survey of Space and Time (LSST, \citealp{LSST..2019}) and the ESA Euclid space mission (\citealp{Euclid..2024}), morphological classification using citizen science project would be impracticable and demands alternative automated methods for speeding up the process.\\

Automated approaches using traditional machine learning or the deep learning algorithms are well-established astronomical research tools used for various astronomical studies. There are several applications of ML techniques in different astrophysical problems like star/galaxy classification, detection of galaxy mergers, finding strong gravitational lenses, predicting the photometric redshift of galaxies, etc. \citet{Banerji2010} utilized the SDSS data and Galaxy Zoo labels for 3 classes to train an Artificial Neural Network (ANN). The trained network is able to reproduce the human classification in GZ project up to the accuracy of 90\%. Deep Convolutional Neural Networks (DCNNs) have significantly improved the process of morphological classification of galaxies because of their ability to learn complex patterns from the raw images. \citet{2015MNRAS.450.1441D} have used DCNN trained on Galaxy Zoo datasets and obtained the morphological classification of 55,000 galaxies. \citet{Abraham..2018} used CNN to produce a catalogue of 25,781 barred galaxies from the SDSS DR13 database. \citet{Prakash..2020} used DCNNs in determining the fundamental parameters governing the dynamical modelling of interacting galaxy pairs. \citet{Sarkar..2023} developed a DCNN to classify spiral galaxies into grand-designs with prominent spiral features and flocculents with fragmented spiral features. \citet{Savchenko..2024} applied their trained DCNN to prepare a catalogue of edge-on galaxies based on the data from the Panoramic Survey Telescope and Rapid Response System (Pan-STARRS, \citealp{panstarrs..2010}). Recently, \citet{Fontirroig..2014} trained a random forest classifier on the lopsided and symmetric galaxies from the IllustrisTNG simulation and achieved an accuracy of $\sim$ 80\%. \\

In this study, we fine-tune a Zoobot (\citealp{Zoobotcode..2023}) model based on \texttt{ConvNeXT\_nano}, a deep learning Python package pre-trained on Galaxy Zoo dataset, for the purpose of identifying lopsided spiral galaxies. The network is trained on the lopsided and symmetric spiral galaxies identified from the SDSS DR18 database. The trained model is then employed to identify a separate larger set of lopsided spiral galaxies from SDSS DR18. We then use this larger sample of lopsided galaxies to study the distribution of their physical properties like specific star formation rate, $g-i$ colour, stellar mass and concentration index. Finally, we address the dependence of the performance of the trained model on photometric parameters of galaxies like redshift and different spiral morphologies. The paper is organized as follows. We describe the selection criteria for the model training dataset in Section \ref{sec:sample} and the DCNN architecture used in this study in Section \ref{sec: architecture}. We present the results in 
Section \ref{sec:result}. We include a discussion in Section \ref{sec: discussion}  and finally present the conclusions in Section \ref{sec:conclusion}.

\section{The Sample} 
\label{sec:sample}

\subsection{Sample Selection}
\label{subsec: selection}
In this study, we select nearly face-on spiral galaxies from the SDSS Data Release 18 (DR 18). SDSS provides both photometric and spectroscopic information for the galaxies (\citealp{Almedia..2023}). To build our sample, we start by selecting all primary photometric objects in SDSS DR18 that are classified as “galaxies” (\texttt{TYPE} flag equal to 3) with the spectroscopic redshifts in the range $0.01 < z < 0.1$ and further, constraining the sample to an extinction-corrected g-band model magnitude \textit{modelMag\_g} $<$ 16 and Petrosian radius (enclosing 90 \% of the flux) $ petroR90\_g \geq \text{3 arcsec}$. This leaves us with a sample of 29,657 galaxies which are sufficiently resolved and also eliminates contaminating foreground objects. The above choice of constraints is motivated with other previous studies aimed at morphological classification (\citealp{Nair..2010, Willett..2013,Abraham_2025}). Next, to extract a subsample of spiral galaxies out of 29,657 galaxies, we utilize the SDSS parameter \textit{fracDeV\_g}. In SDSS pipeline, the surface brightness distribution for each galaxy is fitted using a de Vaucouleurs profile and an exponential profile, and the best-fitting linear combination of both the profile is used to model the distribution. Here, \textit{fracDeV\_g} represents the fraction of luminosity contributed by the de Vaucouleurs profile, and is constrained to lie between 0 and 1. The smaller value of \textit{fracDeV\_g} indicates a disc-dominated profile (late-type) while a higher value represents either a bulged-dominated disc or an elliptical galaxy (early-type). Although various studies have adopted different thresholds to separate between the late-type and early type galaxies, we use $fracDeV\_g \leq 0.5$ (as adopted by \citealp{Shao..2007} in their studies of late type spirals) to select disc-dominated galaxies. This criteria leaves us with 10,528 galaxies which are dominated by exponential profile. We further restrict our sample with the condition $expAB\_g \geq 0.60$ so as to select nearly face-on galaxies because it is difficult to identify lopsidedness for the highly inclined galaxies. The SDSS parameter $expAB\_g$ denotes the axial ratio $B/A$ with $A$ and $B$ as the semi-major and semi-minor axes, respectively. Finally, we are left with a sample of 5,557 nearly face-on galaxies. We download RGB cutouts from the SDSS DR18 in jpeg format with size $256 \times 256$ pixels. \\


Finally, we also consider spiral galaxies which have $fracDeV\_g > 0.5$ with an additional constraints based on the debiased vote fraction from the Galaxy Zoo 2 (GZ2) project for the galaxy to have spiral structure (\textit{t04\_spiral\_a08\_spiral\_debiased}) to be greater than 0.6. These criteria leave us with an additional sample of 2,336 candidates for nearly face-on spirals. Upon visual inspection into the above set of 2,336 galaxies, we extract a reliable set of 1,485 face-on spirals, while remaining (595) were discarded for either not being face-on or being misclassified as spirals based on GZ2 voting criteria. In total, we have a sample of 7,042 (= 5557 + 1485) nearly face-on spiral galaxies. Out of these 7,042, we select a subsample of galaxies for training a Deep Convolutional Neural Network (DCNN) for the task of binary classification (lopsided and symmetric) and the trained DCNN is then utilised for an automatic classification on the remaining samples. Note that although the parameter $fracDeV\_g$ has been used in previous studies to select spiral galaxies, imposing a constraint on $fracDeV\_g$, while yielding a sample largely dominated by spirals, may still admit a small fraction of early-type galaxies. We do not visually inspect the entire sample to exclude such cases. However, in Section \ref{subsec:training_data}, where we visually assess lopsidedness to construct the training sample, we include pure spirals to minimize any contamination from non-spirals in our training set. 


\subsection{Training data}
\label{subsec:training_data}

Since DCNNs are supervised machine learning algorithms, the training data must be labeled. As there is currently no large, publicly available catalog of galaxies hitherto classified as lopsided or symmetric, we begin by preparing a suitable training sample by annotating a subset from the above extracted nearly face-on spirals as either lopsided or symmetric. Visually, lopsidedness can be identified in face-on images when one side of the disk appears more extended than the other or when the geometric center does not coincide with the photometric center. However, visually inspecting a large sample of 7,042 galaxies to identify lopsidedness is both unfeasible and prone to introducing bias. Therefore, we randomly select a smaller, more manageable subset for our training sample. To ensure representative sampling across redshift, we divide the total redshift range ($z = 0.01-0.1$) into four equal intervals, selecting approximately the same number of galaxies ($\sim250$) from each bin. We chose redshift as the criterion for sampling because it is directly related to the effective resolution of the images (later, used for the training). Generally, this approach helps minimize potential biases arising from resolution differences. Table \ref{tab:frac_train} lists the total number of galaxies in the parent sample for each redshift bin, along with the number and fraction (relative to the parent sample in each bin) included in the training set. Next, to preserve the quality of the training data, we exclude some ambiguous cases with uncertain classification labels. This process yields a total of 934 galaxies, comprising 490 classified as lopsided and 444 as symmetric. Although this number is relatively small for conventional machine learning models trained from random weight initialization, it is considered optimal for effectively fine-tuning a domain-specific pretrained model (\citealp{Zoobotcode..2023}). \\
\begin{figure*}[h]
    \centering
    \begin{minipage}[t]{0.48\textwidth} 
        \centering
        \includegraphics[width=\linewidth]{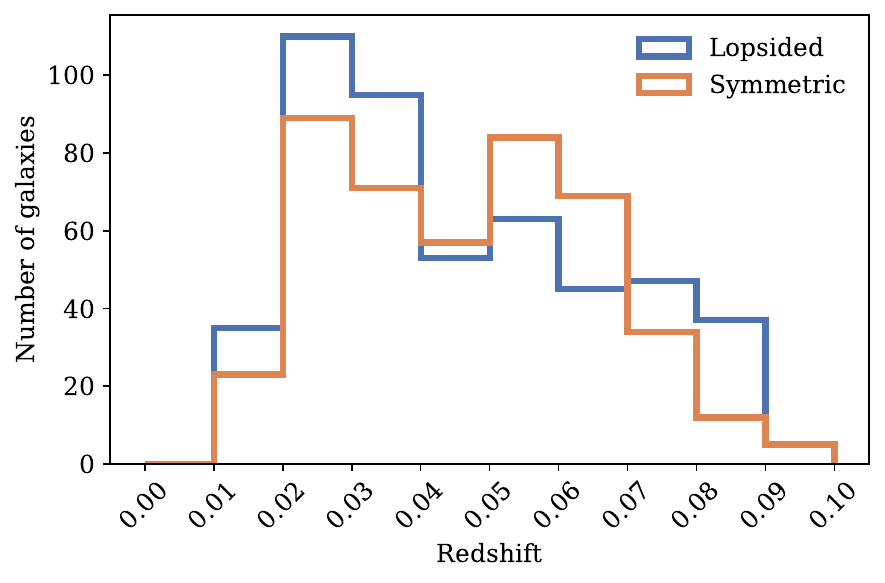}
    \end{minipage}
    \hfill
    \begin{minipage}[t]{0.48\textwidth}  
        \centering
        \includegraphics[width=\linewidth]{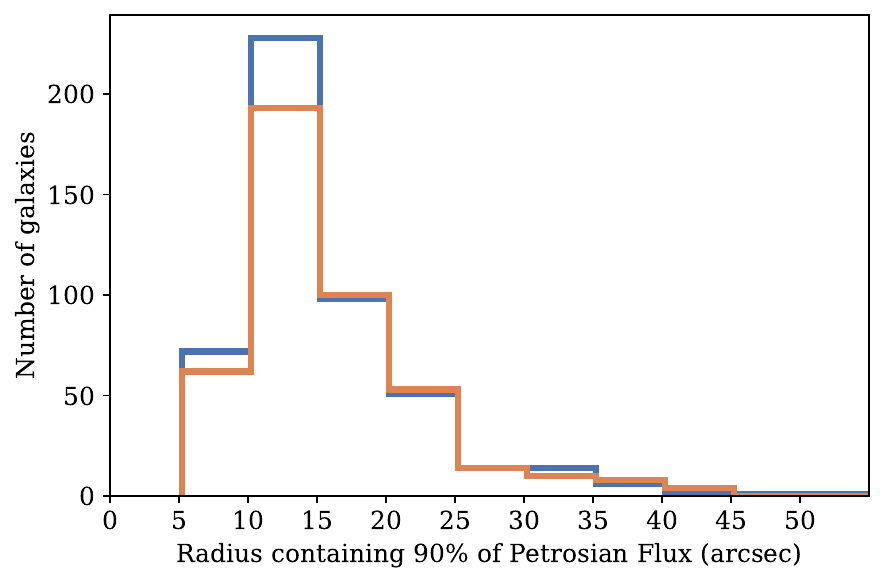}
    \end{minipage}
    \vspace{5mm} 
    \begin{minipage}[t]{0.48\textwidth}  
        \centering
        \includegraphics[width=\linewidth]{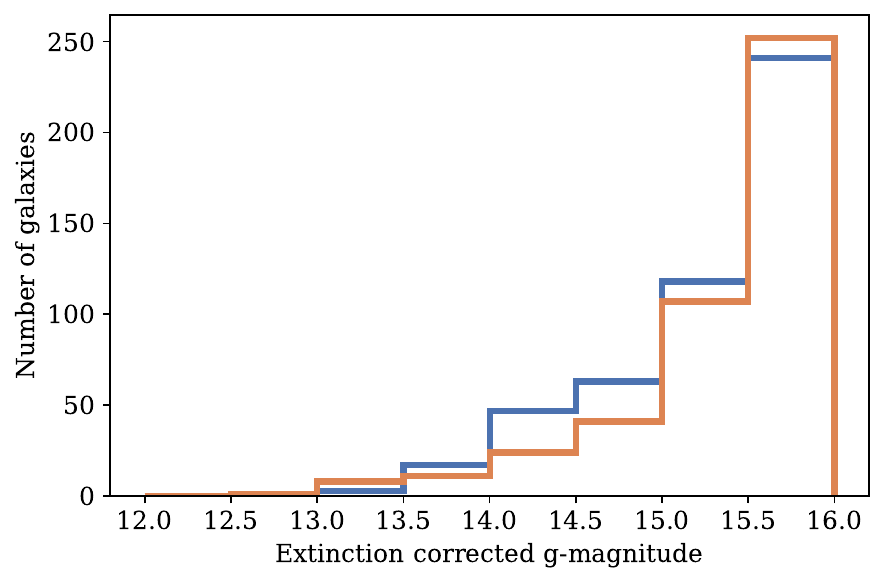}
    \end{minipage}
    \hfill
    \begin{minipage}[t]{0.48\textwidth}  
        \centering
        \includegraphics[width=\linewidth]{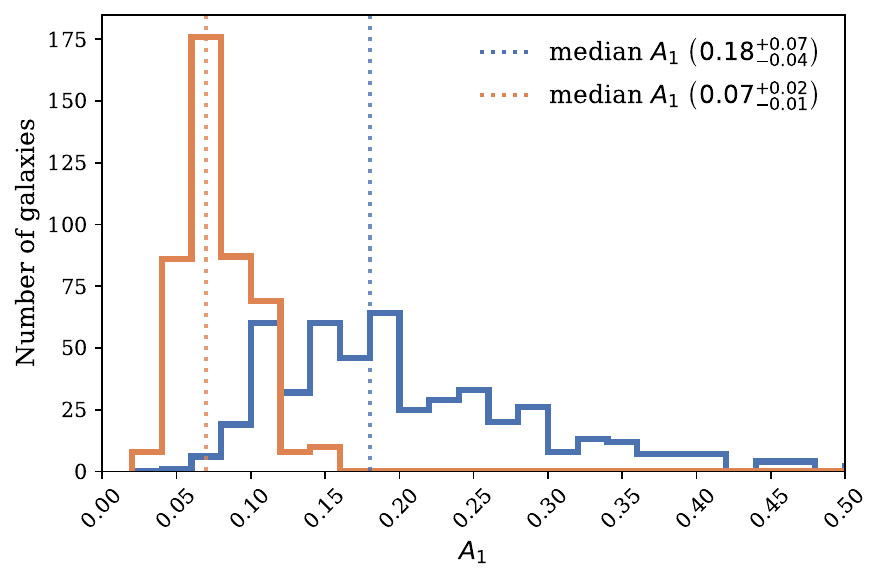}
    \end{minipage}
    \vspace{-5mm}
    \caption{Clockwise from top left: The distributions of redshift, Petrosian radius (enclosing 90 \% of the flux) in the g-band, extinction-corrected g-band model magnitude \textit{modelMag\_g}, and $A_1$ for the samples of lopsided and symmetric spiral galaxies in the training set. A few galaxies were excluded from the Petrosian radius and $A_1$ plots to avoid excessive scaling.}
    \label{fig:hist}
\end{figure*}

\begin{figure*}[t]
    \centering
    \begin{tabular}{cc}
        \includegraphics[width=0.48\textwidth]{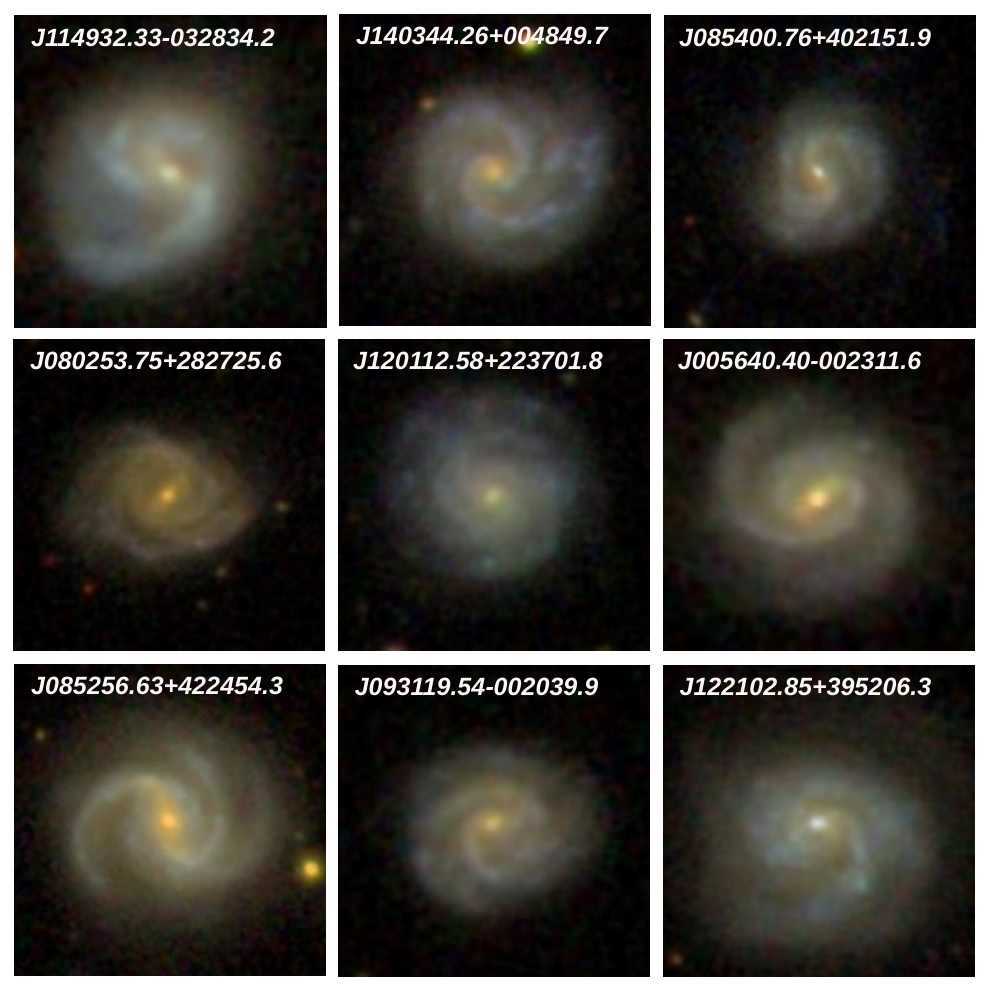} &
        \includegraphics[width=0.48\textwidth]{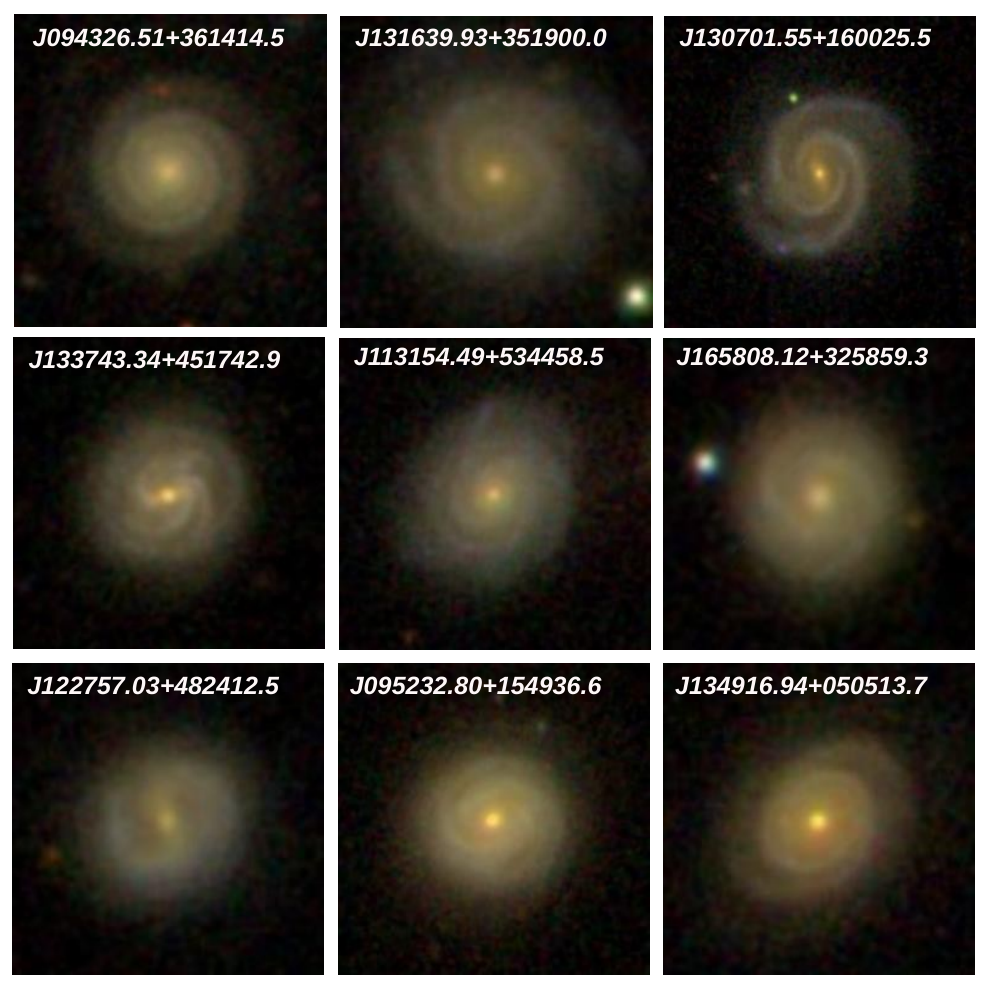} \\
        \text{(a) Lopsided} & \text{(b) Symmetric} 
    \end{tabular}
    \caption{A subset of the lopsided (left) and symmetric (right) galaxies from the SDSS DR18 that are used for the training.}
    \label{fig:train_collage}
\end{figure*}

\begin{table}[h]
\centering
\caption{Number of galaxies in the parent sample and in the derived training set, binned by redshift. The values in parentheses indicate the percentage of the training sample relative to the parent sample in each bin.}
\label{tab:frac_train}
\begin{tabular}{ccc}
\midrule
\textbf{Redshift} & \textbf{Parent sample} & \textbf{Training Sample} \\
\midrule
0.01-0.03 & 2,515 & 257 (10\%)\\ 
0.03-0.05 & 2,776 & 276 (10\%)\\
0.05-0.07 & 1,314 & 261 (20\%)\\ 
0.07-0.1 & 437 & 140 (32\%)\\ 
\midrule
Total & 7,042 & 934 \\
\bottomrule
\end{tabular}
\end{table}

\noindent \textit{Quantifying the degree of lopsidedness:} Besides the visual inspection, we quantify lopsidedness for the training sample. Lopsidedness is quantified by the normalized amplitude of the first mode ($m=1$) in the Fourier decomposition of the surface brightness distribution $I$ of a galaxy (\citealp{Rix..Zaritsky..1995}). To perform Fourier decomposition, we divide a galaxy image into concentric circular annuli, centered at the galaxy centre and perform azimuthal Fourier series decomposition of the surface brightness $I$ in each annulus according to the expression,
{\small
\begin{equation}
I (r_k, \phi)= a_0 (r_k) + \sum^{m_{max}}_{m=1} \Big[ a_{m} (r_k) \cos(m (\phi - \phi_m(r_k)) + b_{m} (r_k) \sin(m (\phi -\phi_m(r_k)) \Big] .
\label{eq:fourierd}
\end{equation}}
where, $a_m(r_k),b_m(r_k) $ represents the amplitude for the mode $m$ corresponding to the $k$th circular annulus (at radius $r_k$) and $\phi_m(r_k)$ denotes the phase. The normalized value of $m=1$ mode amplitude $A_1(r_k)$ is obtained by dividing $\sqrt{a_{1}(r_k)^2 + b_{1}(r_k)^2}$ by the zeroth mode $a_0(r_k)$ (mean surface brightness). $A_1(r_k)$ is the value of lopsidedness for the $k$th circular annulus and when averaged within some chosen radial range within the galaxy gives the value of lopsidedness $A_1$ for the galaxy. \\

Using the \texttt{astroquery.skyview()} module, we download $i$-band (7480 \text{Å}) FITS images ($1000 \times 1000$ pixels) for 934 galaxies from SDSS DR18. We choose \textit{i}-band to trace the old stellar population. However, value of $A_1$ does not vary significantly across the SDSS \textit{g, r,} and \textit{i-}bands (\citealp{Reichard..2009}). Next, sky background are determined using the SEP (\citealp{Barbary..2018}) implementation of the SExtractor package (\citealp{Bertin..1996}) and the mean background is subtracted from each image. The foreground stars are also detected using SEP and the regions were masked by interpolating with the pixel values chosen from nearby regions.  Although the FITS files are downloaded with the target galaxy approximately centered, accurately determining the photometric center is crucial for computing $A_1$. Even a shift of 0.5 pixels can significantly affect the measured $A_1$ value (\citealp{Reichard..2009}). We follow the procedure described in \citet{Reichard..2009} to refine the galaxy centers and compute $A_1$. The center is determined starting from the brightest pixel to be the initial estimate and then further improved by computing the first moment of light in a $3 \times 3$ pixel box centered around the brightest point. Next, we consider the logarithmically spaced radial bins between Petrosian radius containing 50\% and 90\% of the Petrosian flux in the \textit{i}-band ($R_{50}$ and $R_{90}$, respectively), thereby obtaining a series of circular annuli centered on the refined galaxy center. In each annulus, we perform a Fourier decomposition of the surface brightness distribution $I$ by fitting Equation~\ref{eq:fourierd} (up to the fifth order, $m = 5$) using a least-squares approach. The fitting gives us different mode amplitudes $a_m(r_k),b_m(r_k) $ and the phase angle $\phi_m(r_k)$ ($m$ = 0 to 5). The errors associated are derived from the diagonal elements of the covariance matrix. Finally, $A_1$ is computed as the weighted average of the normalized $m=1$ amplitude ($\sqrt{a_{1}(r_k)^2 + b_{1}(r_k)^2}/a_0(r_k)$) between $R_{50}$ and $R_{90}$, where the weights correspond to the the error in each annulus. The median $A_1$ value for lopsided galaxies in the training set is $0.18_{-0.04}^{+0.07}$, while that for symmetric galaxies is $0.07_{-0.01}^{+0.02}$, where the superscripts (subscripts) represent the difference between the third (first) quartile and the median value. Since the computation of $A_1$ is subject to systematic uncertainties arising from several steps in the pre-processing and computation processes (most notably the uncertainty in determining the galaxy’s centre) and because there is no well-established threshold of $A_1$
that separates lopsided from symmetric galaxies, we prefer the annotations obtained from the visual inspection as more reliable labels for training the DCNN. Figure \ref{fig:hist} represents the distribution for redshift, Petrosian radius (enclosing 90 \% of the flux), extinction-corrected g-magnitude and computed $A_1$ for the lopsided and symmetric galaxies in the training set. Figure \ref{fig:train_collage} presents a subset of galaxy images from the training set. The complete dataset used for training the model is publicly accessible on GitHub\footnote{\url{https://github.com/bijusaha-astro/CNN_lopsided}}.

\section{Zoobot}
\label{sec: architecture}
Zoobot is a publicly-available deep-learning based python package by \citet{Zoobotcode..2023}, trained for the morphological classification problems of galaxies. It has been trained on the Galaxy Zoo (GZ) Evo dataset \footnote{GZ Evo:\url{ https://huggingface.co/collections/mwalmsley/galaxy-zoo-evo}}, which contains about $\sim$820k galaxy images from DECaLS/DESI, HST and SDSS. The labels are based on the responses of GZ citizen science projects. Detailed description about the various labels in the GZ decision tree can be found in \citet{GZ..2022}. Being trained to perform diverse classification tasks based on galaxy morphology, Zoobot can easily be adapted (finetuned) to perform any desired new classification task on galaxies with relatively lesser number of labelled images. This makes Zoobot particularly suitable for our problem, because the number of annotated images for lopsided and symmetric galaxies is limited. In contrast, models with randomly initialized weights typically require a much larger training set when trained from scratch. Before training, we split our labelled dataset (comprising 934 galaxies) into training and testing set by the 80:20 ratio. This results in preserving 186 galaxies of both kinds as the test dataset and the remaining 748 galaxies are used for training+validation. 

\subsection{Image Augmentation}
The image dataset comprising 748 samples is divided into training and validation subsets following an 80:20 split, with 80\% allocated for training and 20\% for validation. Some basic image augmentations are applied for training and validation sets to mitigate the limited sample size,
\begin{enumerate}
    \item Rotation in the range -180 \textdegree to 180 \textdegree;
    \item Horizontal and vertical flipping; and
    \item Zoom in or out by 10 per cent.
\end{enumerate}
The details of the total number of images before and after augmentation for the training, validation and testing sample is shown in Table \ref{tab:train_test_count}. No augmentation was applied to the test set. Note that although data augmentation increases the effective training set size (by a factor of 4.5, here), such strong augmentation has the possibility to introduce biases due to repeated use of the same underlying samples. However, the ability of data augmentation to improve model generalization is based on the assumption that the applied augmentations introduce meaningful variations through the above set of geometric transformations.\\

\begin{table}[h]
\centering
\caption{Sample size for training, validation and testing for both the classes combined.}
\vspace{-2mm}
\begin{tabular}{@{}c c c@{}}
\toprule
& \textbf{Before Augmentation} & \textbf{After Augmentation} \\
\midrule
Training & 598 & 2810 \\
Validation & 150 & 930 \\
Testing & \multicolumn{2}{c}{186} \\
\bottomrule
\end{tabular}
\label{tab:train_test_count}
\end{table}

\subsection{Training Zoobot}
\begin{figure}[h]
    \centering
    \begin{minipage}[t]{0.50\textwidth} 
        \centering
        \includegraphics[width=\linewidth]{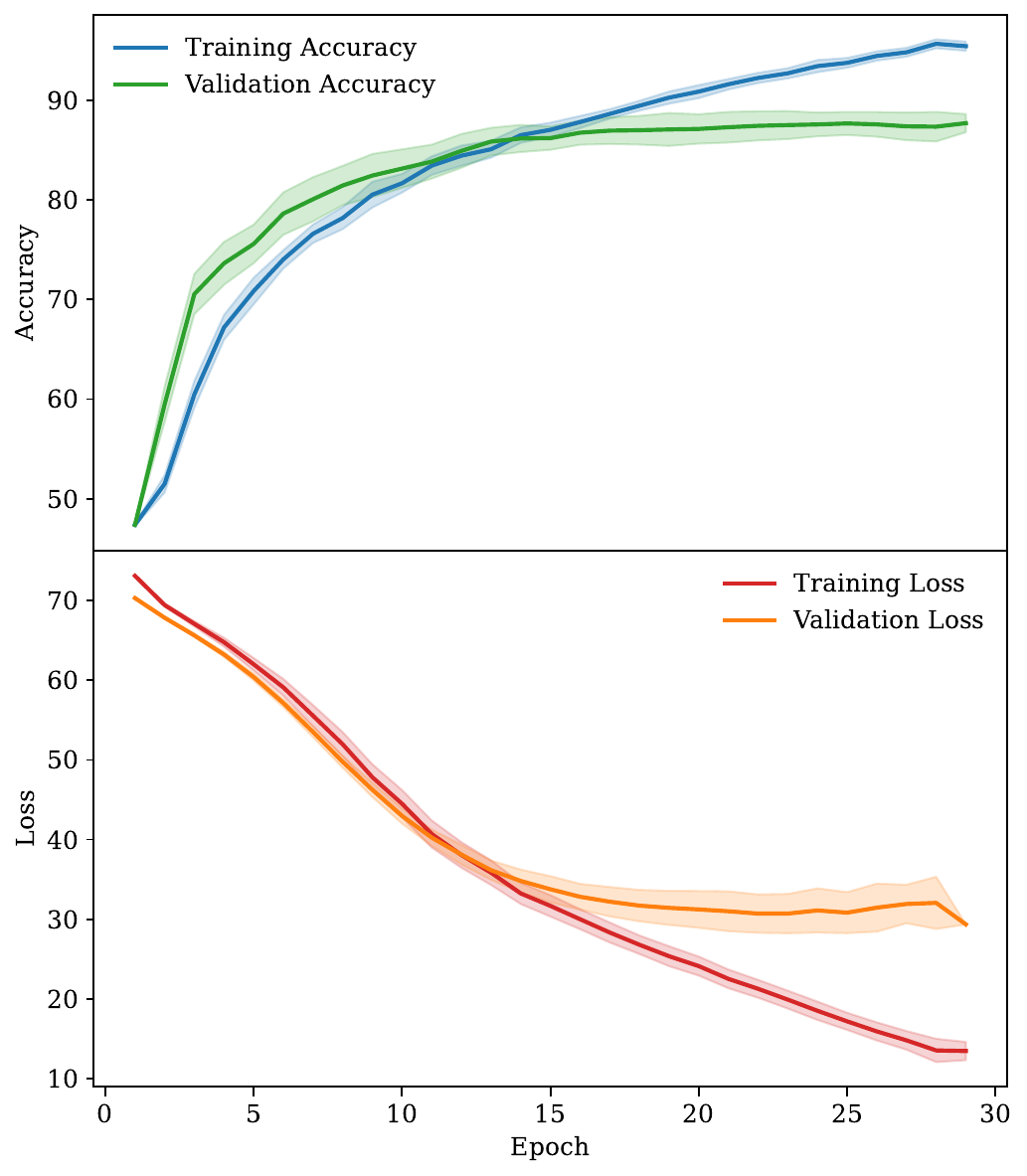}
    \end{minipage}
    \caption{The loss and accuracy of the model as a function of epochs. The solid line represents the mean over 10 independent trials, while the shaded band indicates the $\pm1\sigma$ interval.}
    \label{fig:acc_loss_curve}
\end{figure}

In this study, we employ the PyTorch implementation of Zoobot (\texttt{ZooBotV2}) based on \texttt{ConvNeXT\_nano} model as the backbone architecture. The feature extracting backbone (known as the 'base') with the pre-trained weights is loaded as an encoder from HuggingFace \footnote{\url{https://huggingface.co/collections/mwalmsley/zoobot-encoders-65fa14ae92911b173712b874}} using \texttt{timm}. Next, we add a custom head, which takes the feature maps from the encoder as input. The custom head consists of a global average pooling layer, followed by a fully connected dense layer with 128 neurons, a dropout (of 20\%) and a ReLU activation function. Finally, the output layer with two neurons (one for each class) is added. During training, we perform fine-tuning on the last two blocks of the backbone network (keeping the remaining deeper layers frozen) with a learning rate of $10^{-5}$ and layer decay = 0.5. The loss after each training epoch is calculated using the binary cross-entropy function and the optimization is done using the Adam optimiser with a weight decay = 0.05. The training is performed for a maximum of 50 epochs with a training batch size = 64. However, if the validation loss does not decrease for five consecutive epochs, training is halted, as this indicates that the model is no longer learning and further training could lead to over-fitting. The trained model is then used to make predictions on the test set. We use softmax activation function to convert the class score into probabilities. We conduct 10 independent runs using repeated random splits of data into the training+validation and test sets, i.e., each split was generated randomly, and the training process was repeated for every run. This was done to mitigate sampling bias associated with a single train and ensure that a larger portion of the dataset is included in the test set during evaluation. Figure \ref{fig:acc_loss_curve} presents the mean accuracy/loss curve for the training and validation data across 10 runs and the shaded region represents the corresponding $\pm1\sigma$ interval. To train the models, we used an NVIDIA RTX A2000 12 GB GPU (62 GB RAM; 24 physical CPU cores). Each trial required approximately 20 minutes of training before early stopping was triggered. The training and evaluation of the models was done using PyTorch (\citealp{Pytorch..2019}).

\begin{figure}[h]
    \centering
    \begin{minipage}[t]{0.35\textwidth} 
        \centering
        \includegraphics[width=\linewidth]{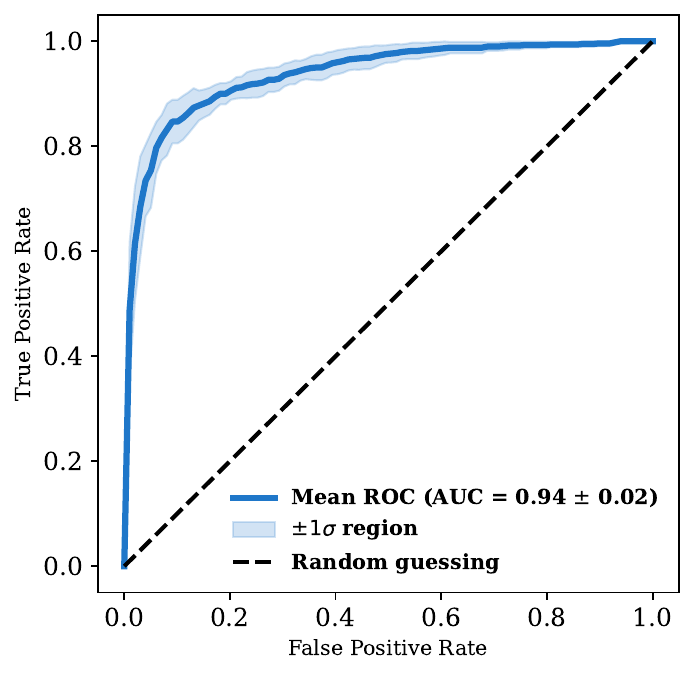} 
    \end{minipage}
    \caption{The Receiver Operating Characteristic (ROC) curve. The solid line shows the mean ROC curve for the 10 independent trials with the shaded region showing the $\pm1\sigma$ interval. The black dashed line indicates a random-guessing classifier with an AUC of 0.5.}
    \label{fig:roc}
\end{figure}

\begin{figure}[h]
    \centering
    \begin{minipage}[t]{0.35\textwidth} 
        \centering
        \includegraphics[width=\linewidth]{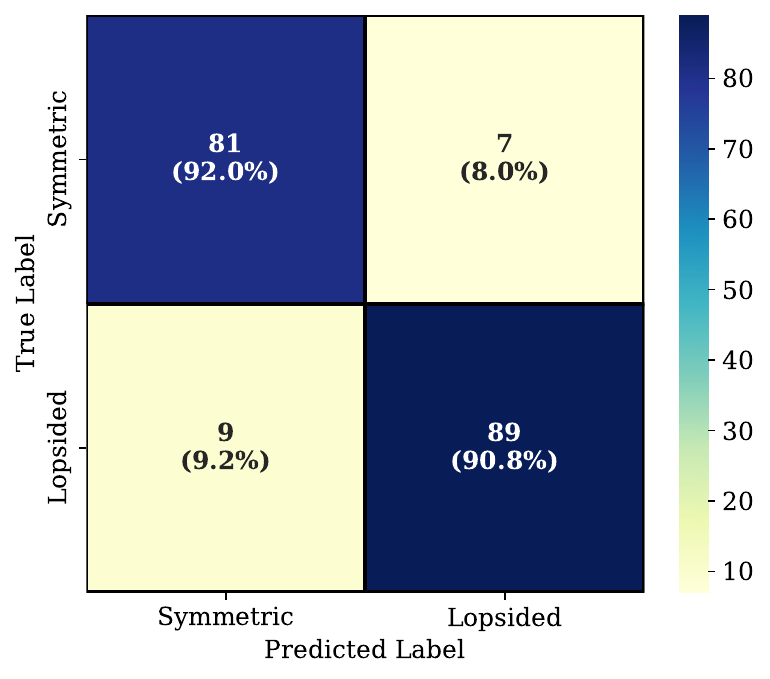} 
    \end{minipage}
    \caption{The confusion matrix representing the correctly predicted and falsely predicted sample in the test set, evaluated using the best-performing model (which has the greatest testing AUC score).}
    \label{fig:confusion}
\end{figure}

\begin{table*}[h]
\centering
\small  
\caption{The classification report for the test dataset, averaged over 10 independent runs with their $\pm1\sigma$ interval}
\begin{tabular}{c c c c}
\toprule
\textbf{Class} & \textbf{Precision} & \textbf{Recall} & \textbf{$f_1$ score} \\
\midrule
Lopsided & $0.89 \pm 0.03$ & $0.87 \pm 0.04$  & $0.88 \pm 0.02$  \\
Symmetric & $0.86 \pm 0.04$ & $0.88 \pm 0.04$ & $0.87 \pm 0.02$ \\
\textbf{AUC} & & \multicolumn{2}{c}{$0.94 \pm 0.02$} \\
\textbf{Accuracy} & & \multicolumn{2}{c}{ $(87 \pm 0.02)$ \%} \\
\bottomrule
\end{tabular}
\normalsize
\label{tab:metric}
\end{table*}

\begin{figure*}[h]
    \centering
    \begin{tabular}{cc}
        \includegraphics[width=0.45\textwidth]{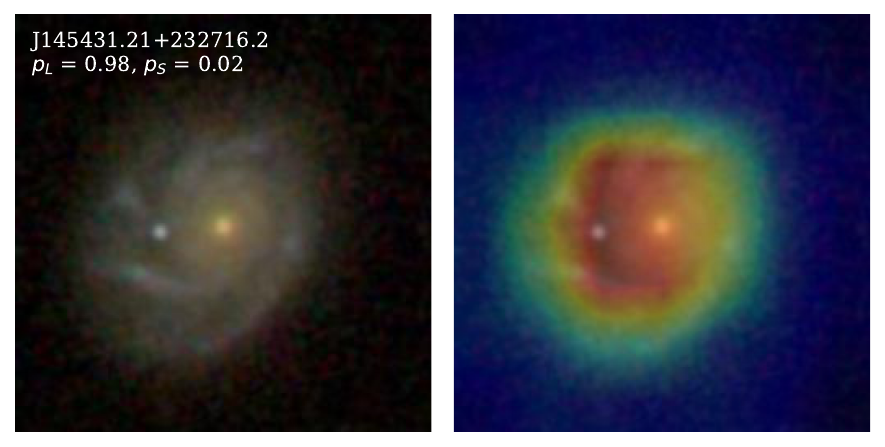} &
        \includegraphics[width=0.45\textwidth]{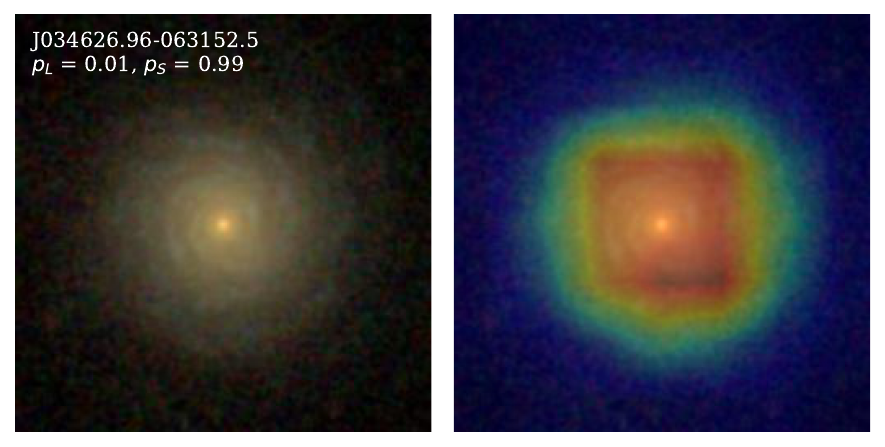} \\

        \includegraphics[width=0.45\textwidth]{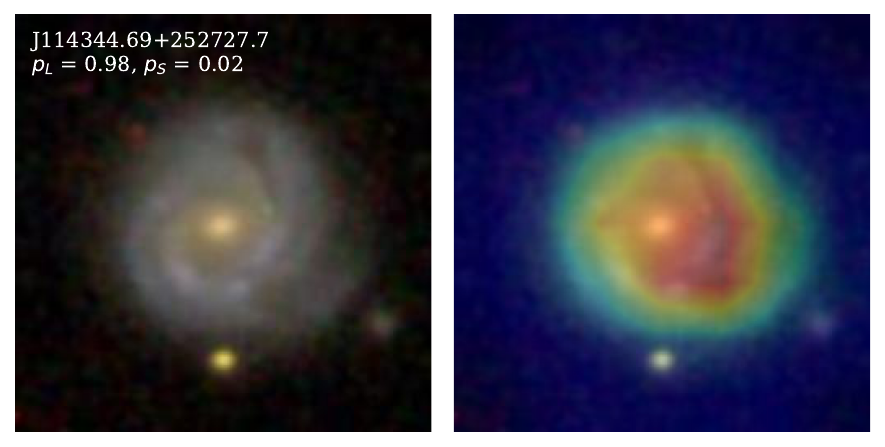} &
        \includegraphics[width=0.45\textwidth]{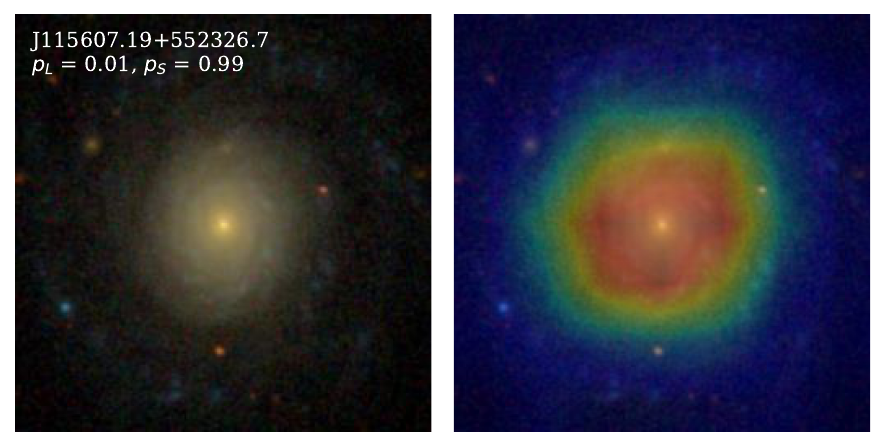} \\
        \text{(a) Lopsided} & \text{(b) Symmetric} \\
       
    \end{tabular}

    \caption{Original images from the test set and the corresponding heatmap from the Grad-CAM analysis. The analysis uses the feature maps from the final convolutional layer (just before the global pooling) of the best-performing model to reveal which spatial regions contributed most strongly to the model’s decision-making process (visualized through a heatmap where yellow/red regions indicate high importance and blue regions correspond to low importance). The left coloumn shows two correctly predicted lopsided galaxies and the right coloumn presents two correctly predicted symmetric galaxies. On the top right corner of each image, the probability with which the model classifies the galaxy as lopsided or symmetric is indicated using the labels $p_L$ and $p_S$, respectively.}
    \label{fig:gradcam}
\end{figure*}

\subsection{Evaluation performance}
\label{subsec:evaluation}
The performance of the trained models on the test set is evaluated using several machine-learning metrics, including precision, recall, $F_1$ score, and the Receiver Operating Characteristic (ROC) curves and its corresponding Area Under the Curve (AUC). The ROC curve plots the True Positive Rate (TPR) against the False Positive Rate (FPR) across all classification thresholds. The AUC provides a measure of how effectively the classifier distinguishes between the two classes. A perfect classifier yields an AUC of 1, whereas a classifier performing random guessing gives an AUC of 0.5. Figure \ref{fig:roc} shows the mean ROC curve obtained from the 10 independent trials, with the shaded region indicating the $\pm1\sigma$ interval. For comparison, we have shown ROC curve for the classifier performing random guessing (AUC = 0.5) as the dashed line. The corresponding mean AUC is $0.94 \pm 0.02$, demonstrating that the classifier performs robustly across different trials. The mean values (with their $\pm1\sigma$ uncertainties) for the other evaluation metrics across the 10 runs are summarised in Table \ref{tab:metric}. Throughout this study, we adopt the default decision threshold of 0.5, i.e., if the predicted probability for a class is greater than 0.5, the sample is assigned to that class. The trained models achieves an accuracy of $(87 \pm 0.02)$ on the test set for the 10 independent trials. Among different trials, we select the model with the highest AUC (AUC = 0.96, accuracy = $91\%$) as the the representative model for subsequent analysis. The confusion matrix for this selected model is shown in Figure \ref{fig:confusion}.  \\

\noindent \textit{Visualisation using Grad-CAM:} One challenge in assessing the reliability of CNN predictions is the limited understanding of what features the model actually learns during the training. As we move deeper into the network, the learned representations reside in increasingly complex latent spaces and therefore become difficult to interpret without any Explainable AI (XAI) techniques. We use one such XAI algorithms known as the Gradient-weighted Class Activation Mapping (Grad-CAM; \citealp{gradcam..2016}) to visualize the regions of the galaxy images that most strongly influence the model’s predictions. Grad-CAM computes the gradients of the target class score with respect to the feature maps of the final convolutional layer. These gradients are then spatially averaged to obtain importance weights, which quantifies the contribution of each feature map to the target class. Finally, linear combination of the feature maps, weighted by the importance are taken to highlight the regions most relevant for the model prediction. Figure \ref{fig:gradcam} shows examples of galaxy images alongside their corresponding Grad-CAM heatmaps overlaid on the original images. The left coloumn of Figure \ref{fig:gradcam} shows two correctly predicted lopsided galaxies, while the right coloumn presents two correctly predicted symmetric galaxies. In the heatmaps, red regions indicate areas of high importance, followed by yellow and green, while blue contribute little to the prediction. We observe that when the model correctly predicts a galaxy as lopsided, the red regions in the heatmap often tend to be heavily concentrated offset from the centre. In contrast, for symmetric galaxies, the heatmap often appears more uniformly distributed throughout the galaxy image. This can be understood as follows: since symmetric galaxies lack distinct asymmetric features, the model assigns roughly equal importance to all regions, resulting in an almost uniform red distribution (right coloumn of Figure \ref{fig:gradcam}) in the heat map. In a few cases, however, the heatmaps for symmetric predictions display small patches of high-importance regions that extend outside the galaxy.

\section{Results}
\label{sec:result}
\begin{figure*}[p]
    \centering

    \begin{minipage}{0.9\textwidth}
        \centering
        \includegraphics[width=\linewidth]{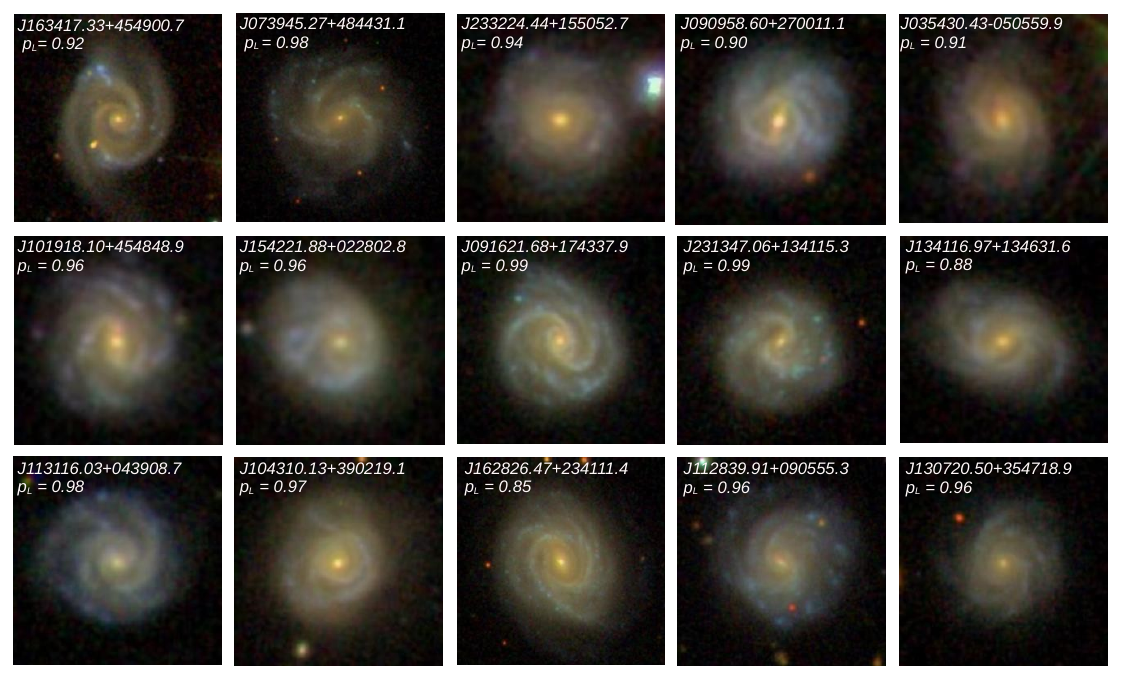}
        \subcaption*{(a) Lopsided}
    \end{minipage}\\[2mm]

    \begin{minipage}{0.9\textwidth}
        \centering
        \includegraphics[width=\linewidth]{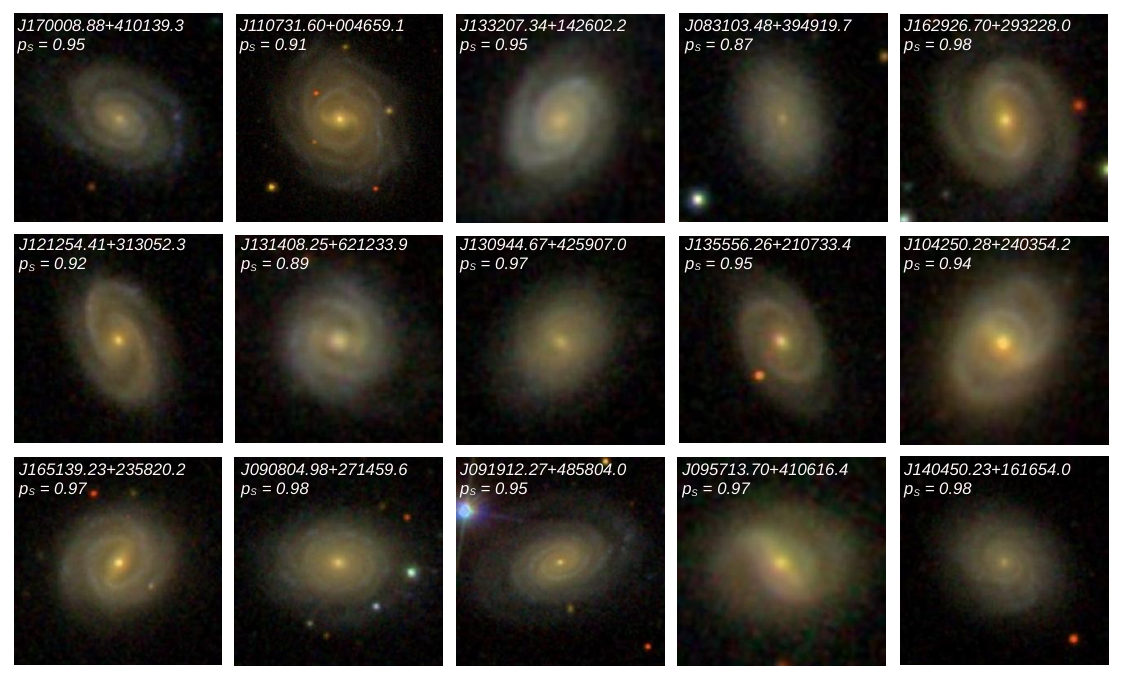}
        \subcaption*{(b) Symmetric}
    \end{minipage}
    \caption{A subset of the (a) lopsided (top row) and (b) symmetric galaxies (bottom row) predicted from SDSS DR18 using the best-performing model. The prediction probability corresponding to the predicted label is shown at the top-right corner of each image.}
    \label{fig:prediction}
\end{figure*}

\begin{figure}[h]
    \centering
    \begin{minipage}[t]{0.50\textwidth} 
        \centering
        \includegraphics[width=\linewidth]{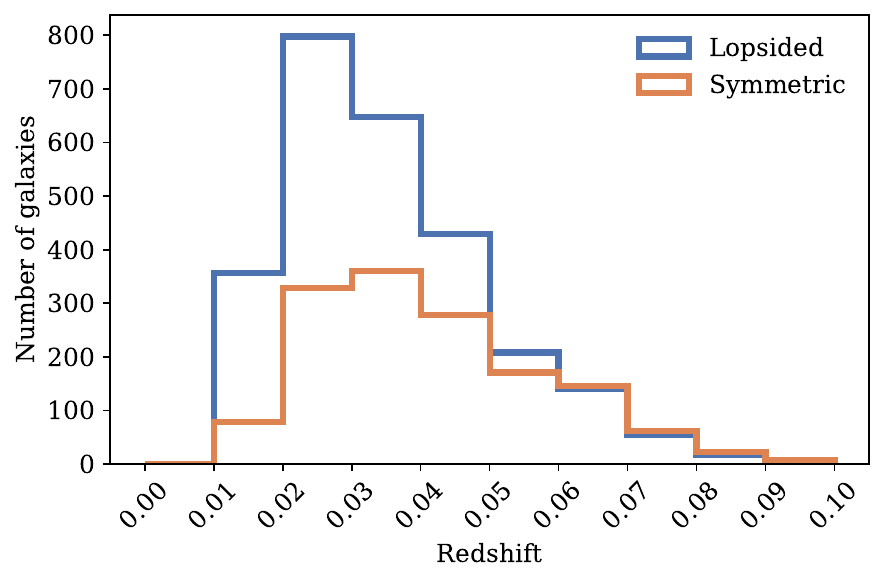}
    \end{minipage}
    \caption{Redshift distributions for the newly predicted galaxy samples, comprising 2,658 lopsided and 1,455 symmetric systems with prediction probability $P_{\mathrm{pred}} \geq 0.85$.}
    \label{fig:redshift_pred}
\end{figure}

We use the best-performing model for automatic predictions on the remaining 6,108 galaxies (as mentioned in Section \ref{subsec: selection}). Based on the default prediction probability threshold $P_{pred}=0.5$, 3,679 (out of 6,108) galaxies are predicted as lopsided and the remaining 2,429 are predicted as symmetric. To ensure a reliable final sample for the statistical study of the physical properties, we choose a threshold of $P_{pred}=0.85$ to represent a high-confidence sample, consisting of 2,658 lopsided and 1,455 symmetric galaxies. We present a subset of the newly predicted lopsided and symmetric galaxies sample in Figure \ref{fig:prediction}. The distribution of redshift for the newly predicted galaxies in the high-confidence sample ($P_{pred} \geq 0.85$) is shown in Figure \ref{fig:redshift_pred}. The details of the newly predicted sample along with the prediction probability for the predicted class are publicly accessible through our GitHub repository\footnote{\url{https://github.com/bijusaha-astro/CNN_lopsided}} . \\

\noindent \textit{Physical properties: } We choose the galaxies predicted with $P_{pred} \geq 0.85$ to carry out a statistical study of their physical properties. We use the specific star-formation rate ($sSFR$) and stellar mass ($M_*$) from the SDSS \textit{StellarMassFSPSGranWideDust} table and Petrosian \textit{g - i} colour from the SDSS \textit{PhotoObj} table. We also use the Petrosian radii to obtain the concentration index $C_{i}$.  $C_{i} = R_{90}/R_{50}$ is used as a proxy for the Hubble type, where a higher value of $C_{i}$ corresponds to an early-type galaxy (\citealp{strateva..2001}, \citealp{Shimasaku..2001}).\\

\clearpage
\clearpage

\begin{figure*}[ht]
    \centering
    \begin{minipage}[t]{0.48\textwidth} 
        \centering
        \includegraphics[width=\linewidth]{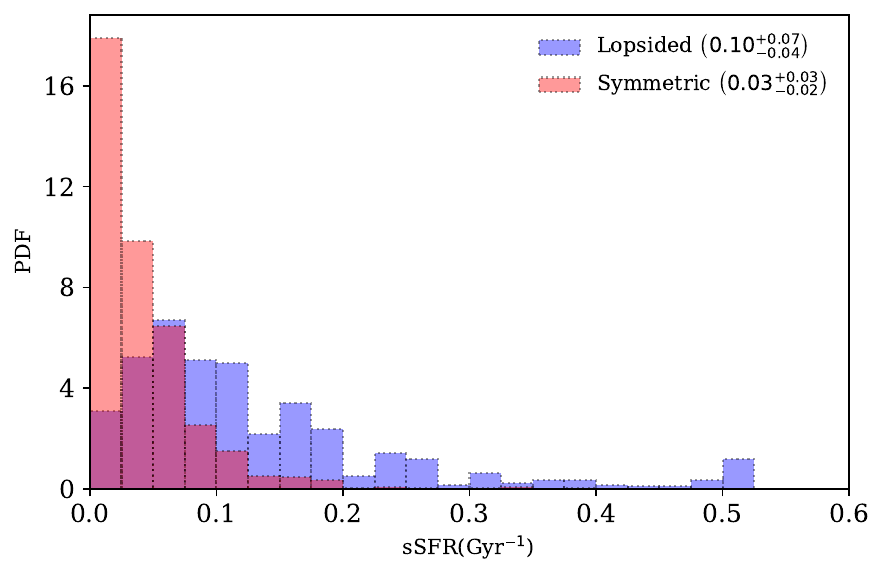}
    \end{minipage}
    \hfill
    \begin{minipage}[t]{0.48\textwidth}  
        \centering
        \includegraphics[width=\linewidth]{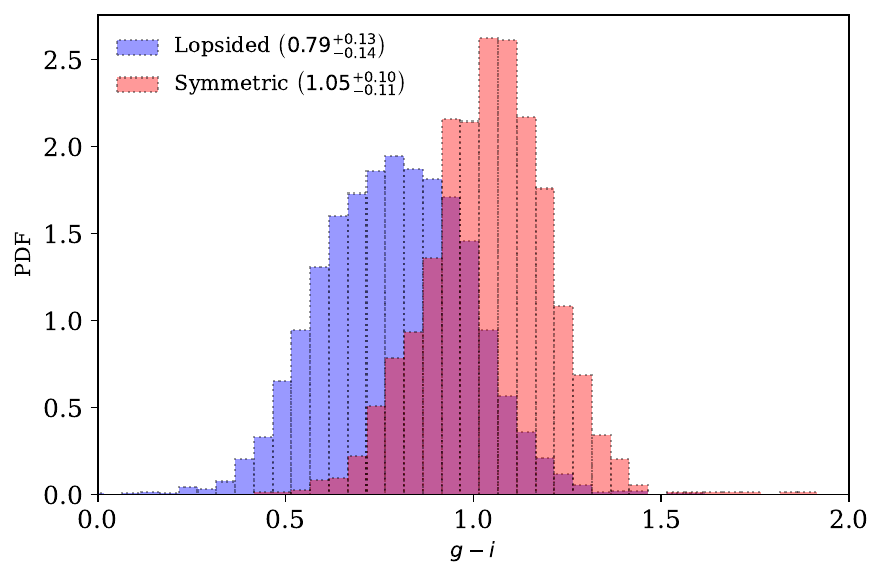}
    \end{minipage}
    \vspace{5mm} 
    \begin{minipage}[t]{0.48\textwidth}  
        \centering
        \includegraphics[width=\linewidth]{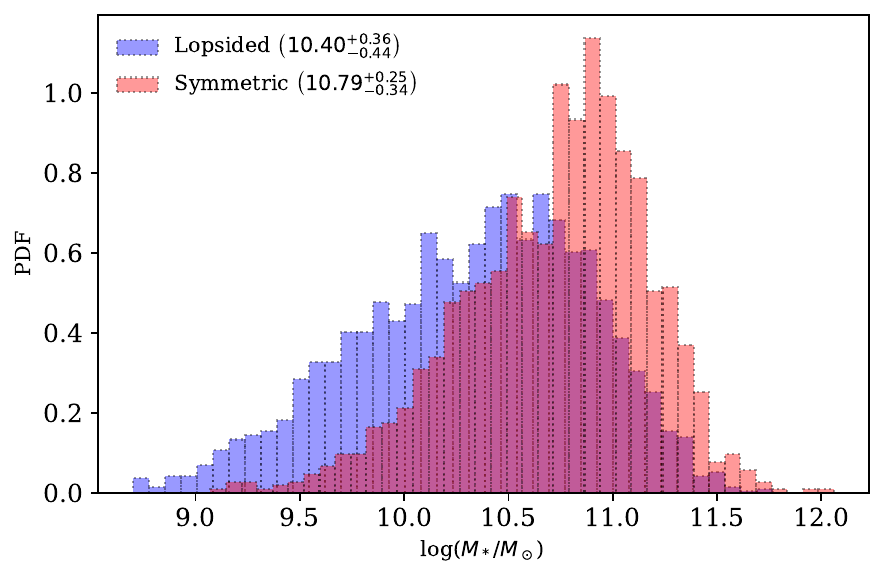}
    \end{minipage}
    \hfill
    \begin{minipage}[t]{0.48\textwidth}  
        \centering
        \includegraphics[width=\linewidth]{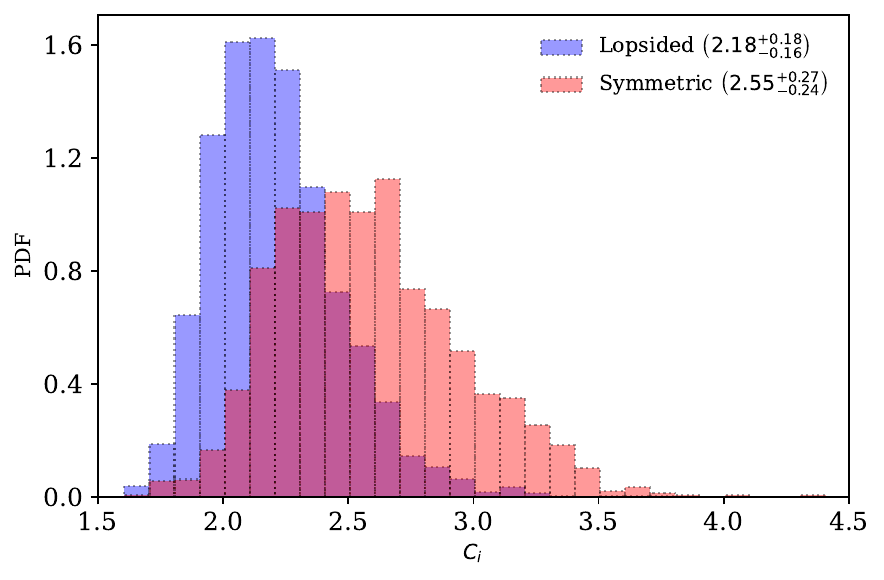}
    \end{minipage}
    \vspace{-5mm}
    \caption{Clockwise from top left: Probability density function (PDF) of specific star formation rate (sSFR in Gyr$^{-1}$), $g - i$ colour, log-stellar mass ($\log(M_*/M_\odot)$) and concentration index ($C_i = R_{90}/R_{50}$) for the newly predicted samples of lopsided and symmetric spiral galaxies.}
    \label{fig:sSFRandcolour}
\end{figure*}

In Fig: \ref{fig:sSFRandcolour}, we show the Probability Density Functions (PDFs) of the specific star formation rate (sSFR in Gyr$^{-1}$), $g - i$ colour, log-stellar mass ($\log(M_*/M_\odot)$) and concentration index ($C_i = R_{90}/R_{50}$) for the newly predicted samples of lopsided and symmetric spiral galaxies. The median values of the above physical properties of the newly classified sample are presented in Table \ref{tab:median_prop}. Additionally, to mitigate any redshift-driven bias arising from the under-representation of symmetric galaxies in our predicted sample at low redshift ($z < 0.05$, see Figure \ref{fig:redshift_pred}), we adopt a stratified subsampling approach to ensure equal representation of both categories across the entire redshift range $0.01 \leq z \leq 0.1$. In the higher-redshift range ($0.05 \leq z \leq 0.1$), where the lopsided and symmetric populations have comparable numbers, no subsampling is applied and all galaxies are retained. In contrast, the redshift range $0.01 \leq z < 0.05$ shows imbalance between the two populations. We therefore divide this range into bins of width $\Delta z = 0.01$ and, within each bin, randomly subsample the lopsided population so that its size matches the number of symmetric galaxies (the minority class). This procedure yields a balanced redshift distributions for the lopsided and symmetric samples and the median value of a given physical property is computed. To assess the stability of the results against random selection, the subsampling procedure is repeated 100 times. The mean of the median values obtained across these 100 realizations is reported in parentheses in Table~\ref{tab:median_prop}. The standard deviation of the median values across different realizations is found to be negligible and is therefore not explicitly reported.\\

We find that the lopsided galaxies in our predicted samples are relatively high star-forming, bluer, low-concentration (late-type), low-mass galaxies. This is consistent with the results from a few observational and simulation studies (\citealp{Zaritsky..1997}, \citealp{Reichard..2009}, \citealp{lokas..2022}). The studies suggest a plausible correlation between lopsidedness in the galaxy and the star formation rate and hence, an excessive blue luminosity in the galaxy. Lopsidedness in the galaxy arises from the tidal interactions and merger events, which also leads to the inflow of gas in the galaxy and a subsequent increase in the star formation rate. While the origin of lopsidedness typically involves tidal interactions and minor mergers, strong lopsided features are also observed in field galaxies where galaxy interactions are negligible. \citet{Bournaud..2005} argued that asymmetric gas accretion which leads to long-lived lopsidedness in field galaxies, can also result in increased recent star formation activity. The PDF for the log-stellar mass $\log(M_*/M_\odot)$ suggests that lopsided galaxies are low-mass galaxies. This observation is in line with earlier findings (\citealp{Reichard..2009}, \citealp{VL..2023}). The median value of $C_{i}$ for the lopsided sample is 2.18, while it is 2.55 for the symmetric sample. \citet{Kauffmann..2003} showed that $C_i=2.6$ can be considered as the approximate threshold separating early-type and late-type galaxies, which suggests that lopsided galaxies generally dominate the late-type population. 

\begin{table}[h]
\centering
\caption{Physical properties for the newly predicted sample.}
\label{tab:median_prop}
\begin{tabular}{ccc}
\midrule
\textbf{Physical properties} & \textbf{Lopsided} & \textbf{Symmetric} \\
\midrule
sSFR (in Gyr$^{-1}$) & $0.10_{-0.04}^{+0.07}$  (0.10) & $0.03_{-0.02}^{+0.03}$ \\ [0.3cm]
$g - i$ colour & $0.79_{-0.14}^{+0.13}$ (0.83) & $1.05_{-0.11}^{+0.10}$ \\ [0.3cm]
$\log(M_*/M_\odot)$ & $10.40_{-0.44}^{+0.36}$ (10.55) & $10.79_{-0.34}^{+0.25}$ \\ [0.3cm] 
$C_i$ & $2.18_{-0.16}^{+0.18}$ (2.20) & $2.55_{-0.24}^{+0.27}$ \\ 
\bottomrule
\end{tabular}
\vspace{2mm}
\begin{tablenotes}
\footnotesize
\item \textbf{Note.} The bracketed entries represent the median values for the subsample of lopsided galaxies, obtained by randomly selecting (within each redshift bin $\Delta z = 0.01$) the same number of lopsided galaxies as there are symmetric galaxies to correct for the under-representation of symmetric systems at low redshift ($z < 0.05$). The reported values are averaged over 100 runs of random sampling.
\end{tablenotes}
\end{table}

\section{Discussion}
\label{sec: discussion}

\begin{figure}[h]
    \centering
    \begin{minipage}{0.45\textwidth}
        \centering
        \includegraphics[width=\linewidth]{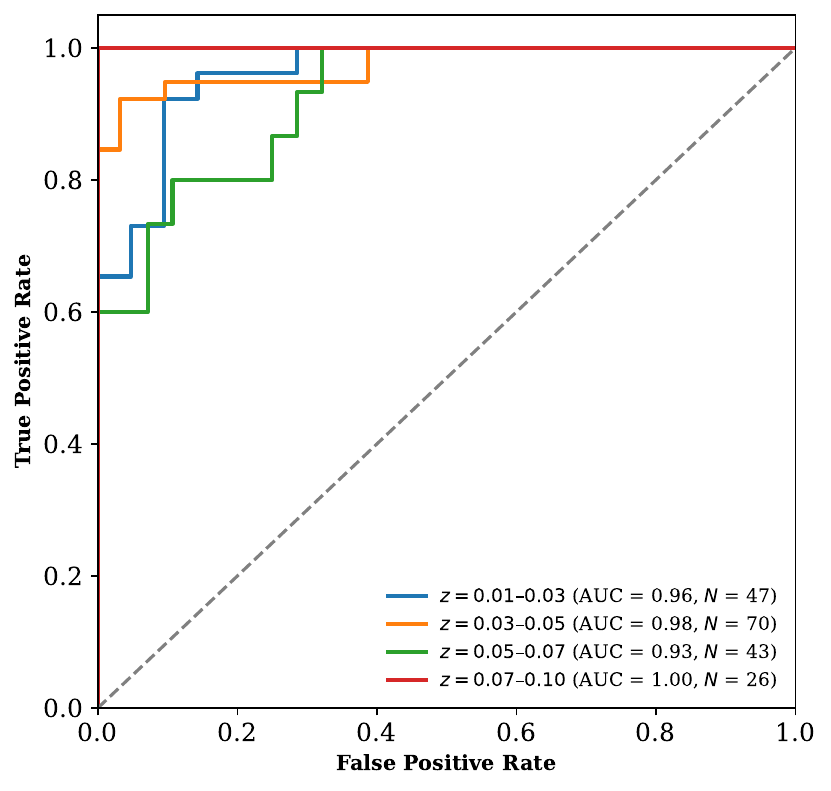}
        \subcaption{}
        \label{fig:auc_a}
    \end{minipage}\\[2mm]
    \begin{minipage}{0.45\textwidth}
        \centering
        \includegraphics[width=\linewidth]{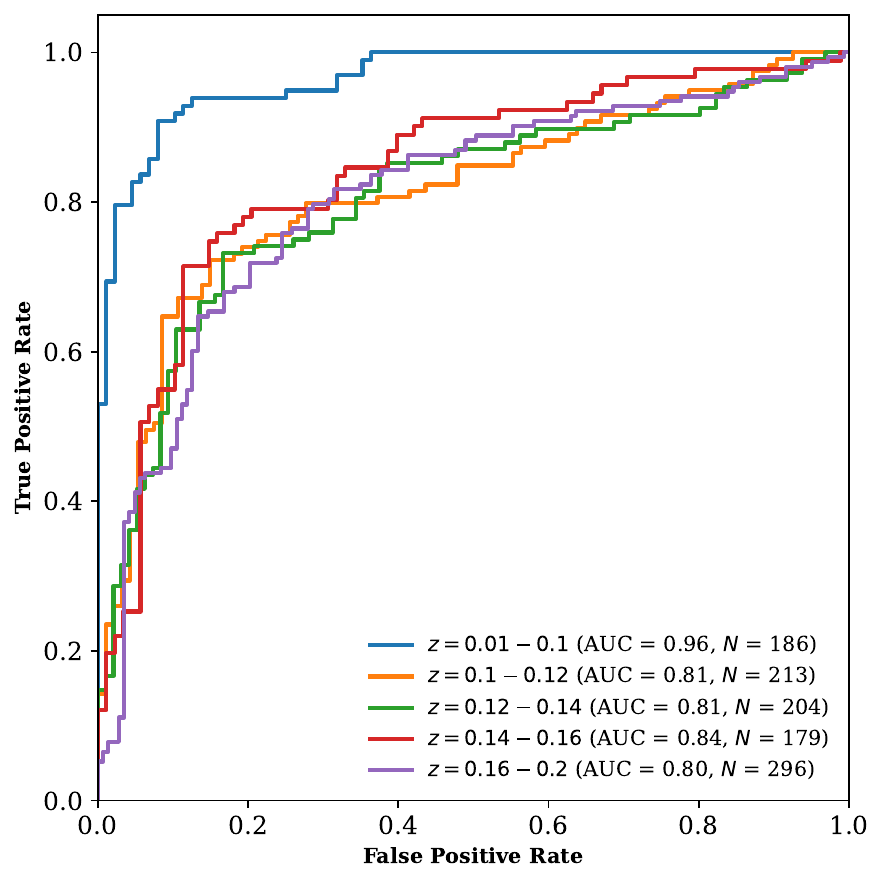}
        \subcaption{}
        \label{fig:auc_b}
    \end{minipage}
    \caption{ROC curves for galaxies at different redshift bins, colour-coded as shown in the legend, with each label also listing the corresponding AUC score and the number of galaxies in the bin $N$. Top Plot (a) shows the ROC curves for different redshift bins within the low redshift range $z = [0.01, 0.1]$ and the bottom plot (b) shows the ROC curves for different redshift bins within the high redshift range $z = [0.1, 0.2]$, along with the ROC curve for the complete low redshift range $z = [0.01, 0.1]$ shown for the comparison. }
    \label{fig:auc}
\end{figure}

\noindent \textit{Dependence of model prediction on Redshift:} In this study, we have restricted ourselves to bright, low redshift galaxies because the training sample is built through visual classification, and during this process we observe that redshift (connected to image resolution) is a key factor in determining whether a galaxy appears lopsided or symmetric. As redshift increases, the contrast between the brighter and dimmer sides of a galaxy diminishes, causing intrinsically lopsided galaxies to appear more symmetric (\citealp{Reichard..2009}), which may introduce bias in labelling. To assess the dependence of model performance on redshift, we divide the test sample in different redshift bins and plot the ROC curve to visualize the model’s performance in each bin across all classification thresholds. The test set is split into four redshift bins: (0.01, 0.03], (0.03, 0.05], (0.05, 0.07], and (0.07, 0.10]. The redshift bins are chosen such that we have approximately equal samples from the test set for each bin. Figure \ref{fig:auc_a} shows the ROC curves for the different redshift bins over the redshift range $z=[0.01,0.1]$ , with the corresponding AUC scores and number of samples $N$ for the particular bin indicated in the legend. Comparing the AUC scores allows us to assess the ability of the model to distinguish between the two classes for different redshift bins. We observe that the model performs fairly well for all redshift bins with AUC > 0.9 for all the redshift bins. However, for the last redshift bin $z = (0.07, 0.01]$, the sample size is small enough for a fair comparison.\\

As a robustness check, we investigate how well the model (which is trained on low-$z$ samples; $z=0.01 - 0.1$) performs at relatively higher $z$ samples. To obtain a representative high-$z$ spiral galaxies from SDSS DR18, we consider the following selection criteria: (i) $0.1 < z <0.2$, (ii) extinction-corrected g-band model magnitude \textit{modelMag\_g} $< 18$, (iii) Petrosian radius $ petroR90\_g \geq \text{6 arcsec}$, (iv) $fracDeV\_g \leq 0.5$, and (v) $expAB\_g \geq 0.60$. These cuts yield 14,312 face-on spiral galaxies for the high-$z$ analysis. To select a representative test sample, we sample from different redshift bins. We divide the redshift interval into five equal-width bins. However, due to sparse sampling and reduced visual certainty at the upper end of the redshift range, we merge the last two bins, resulting in the final redshift binning: (0.10, 0.12], (0.12, 0.14], (0.14, 0.16], and (0.16, 0.20]. From each bin, we randomly select approximately 200 galaxies and visually classify them as lopsided or symmetric to construct a test sample. We then apply our best-performing model to this visually classified set and compute the ROC–AUC score for each redshift bin. Figure \ref{fig:auc_b} presents the ROC curves for different redshift bins within the high-redshift range $z=[0.1,0.2]$. For comparison, the ROC curve for the complete low redshift range $z = [0.01, 0.1]$ is also shown (AUC = 0.96). The model achieves an average AUC score of 0.82 across the high-$z$ bins. Although the AUC score decreases from 0.96 to 0.82 when moving to higher redshift, the model still shows a reasonable ability to predict correctly at higher redshift. It is also important to note that visual classification becomes challenging at higher redshift, introducing additional bias that may affect the evaluation. \\

\begin{figure}[h]
    \centering
    \begin{minipage}{0.45\textwidth}
        \centering
        \includegraphics[width=\linewidth]{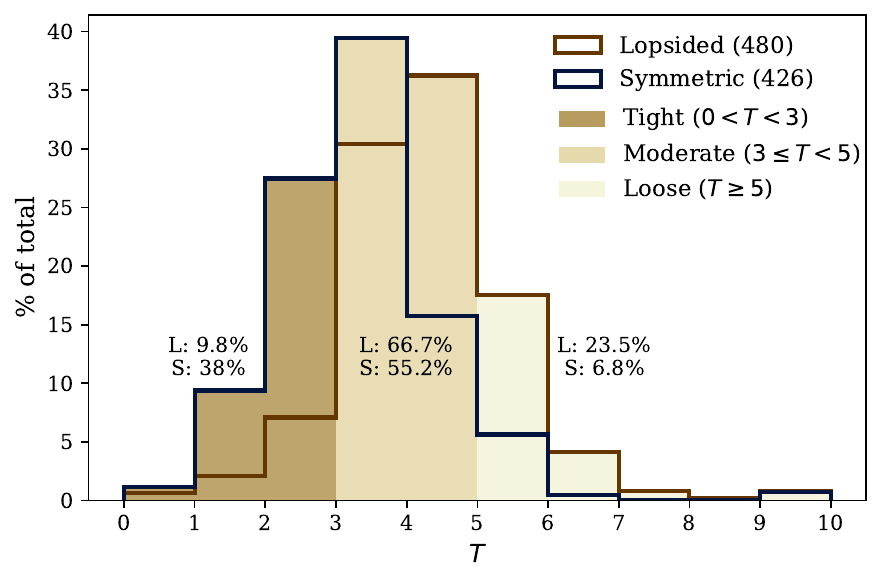}
        \subcaption{Training Set}
        \label{fig:T_type_a}
    \end{minipage}\\[2mm]
    \begin{minipage}{0.45\textwidth}
        \centering
        \includegraphics[width=\linewidth]{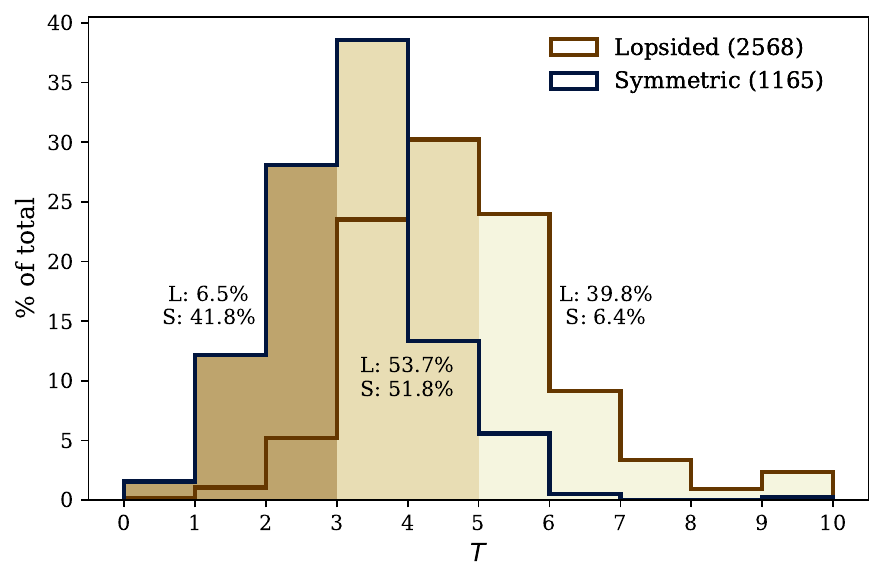}
        \subcaption{Prediction Set ($P_{pred} \geq 0.85$)}
        \label{fig:T_type_b}
    \end{minipage}
    \caption{Percentage of various spiral morphologies based on the tightness of the spiral arm winding. The bracketed number indicates the total number in each set.  Top Plot (a) shows the training set and bottom plot (b) shows the prediction set. The annotated text in the figure shows the percentage of samples for the three different windings.}
    \label{fig:T_type}
\end{figure}

\noindent \textit{Dependence of model prediction on spiral morphology:} In Section \ref{sec:result}, we have shown that lopsidedness is more common in late-type spirals (low $C_i$) with loosely wound arms, whereas symmetric galaxies are more frequently found in early-type spirals (high $C_i$) with tightly wound arms. Although our goal is to classify galaxies as lopsided or symmetric, it is challenging to completely disentangle this dependence on spiral-arm tightness, as it is intrinsically linked to galaxy morphology. One approach to mitigate this is to ensure that the training sample includes sufficient representation of the minor cases, i.e., symmetric galaxies with loose winding and lopsided galaxies with tight winding. To confirm this, we collect the de Vaucouleurs numerical stage indices ($T$) from the HyperLeda database (\citealp{Makarov..2014}), which allow us to identify spiral galaxies based on the tightness of their arm winding. The $T$ values retrieved from the HyperLeda database sometimes show discrepancies with visual morphology; in particular, some early-type spiral galaxies with clearly visible spiral arms are assigned negative $T$ values. As our study focuses on spiral galaxies, we ensure that the training set (constructed through visual classification) contains only pure spirals. However, it is not feasible to visually inspect the remaining larger sample, which may therefore include a small fraction of non-spiral galaxies. Figure~\ref{fig:T_type} presents the distribution of spiral-arm morphologies for both the training set and the high-confidence predicted sample, where the y-axis represents the percentage of the total sample in each bin. In Figure~\ref{fig:T_type_a}, we exclude 18 symmetric and 10 lopsided galaxies from the training set for which the $T$ values are either unavailable or unreliable. In Figure~\ref{fig:T_type_b}, we include only galaxies with $T > 0$ from the predicted sample to focus on the relative populations of different spiral-arm morphologies. In terms of winding, we divide the $T$ values as follows: (a) tight winding (Sa, Sab): $0 < T < 3$ ; (b) moderate winding (Sb, Sbc): $3\leq T < 5$ ; and (c) loose winding (Sc, Scd, Sd, Sm): $T \geq 5 $. Comparing the relative fractions across winding categories, we find that both lopsided and symmetric galaxies (in both the training and prediction sets) are predominantly moderately wound, followed by loosely wound spirals for lopsided galaxies and tightly wound spirals for symmetric ones. Importantly, the presence of minor cases, i.e., tightly wound lopsided galaxies and loosely wound symmetric galaxies ensures that the model prediction are not primarily driven by driven by tightness of spiral arms. Moreover, the GradCAM analysis in Section \ref{subsec:evaluation} further demonstrates that the model successfully captures the relevant morphological features necessary to distinguish between lopsided and symmetric galaxies. 

\noindent \textit{Human Labelling:} Visual inspection was the only preferred method to determine the labels for the training sample. Even though visual inspection of galaxies to identify subtle features like lopsidedness is time-consuming and requires careful attention, especially in the case of galaxies that lie between the most symmetric and the most lopsided. Excluding those galaxies would result in a significant decline in sample size and also restrict the ability of the model to generalize to a larger sample. Owing to the extreme sensitivity of $A_1 $ to various factors (for instance, center determination), careful visual inspection becomes the only reliable method. To minimize biases introduced during the visual-inspection process, we selected a manageable number of galaxies, repeated the inspection procedure with increased scrutiny, and excluded only those cases for which the classification remained uncertain.

\section{Conclusion}
\label{sec:conclusion}
\noindent We fine-tune a Zoobot model to identify lopsided spiral galaxies. We select 7,042 nearly face-on spiral galaxies from SDSS DR18 over the redshift range $0.01 \leq z \leq 0.1$, with extinction-corrected g-band model magnitude \textit{modelMag\_g} $<$ 16 and Petrosian radius (enclosing 90 \% of the flux) $ petroR90\_g \geq \text{3 arcsec}$. Due to the non-availability of a sufficiently large sample of galaxies previously classified as lopsided or symmetric, we construct our training set through careful visual inspection. To make the visual classification manageable, we focus on a relatively smaller subset randomly sampled from the above parent sample, binned across different redshift intervals, resulting in a training sample of 934 galaxies (490 lopsided and 444 symmetric). Next, we fine-tune the Zoobot model based on the \texttt{ConvNext\_nano} architecture, achieving a testing accuracy of $(87 \pm 0.02)\%$, averaged over 10 independent trials. We then use the best-performing model (highest AUC score = 0.96, accuracy = 91\%) to make predictions for the remaining 6,108 galaxies, identifying 3,679 lopsided and 2,429 symmetric galaxies. Using the subset of 2,658 lopsided and 1,455 symmetric galaxies predicted with high probability ($P_{pred} \geq 0.85$), we study their physical properties. The lopsided galaxies have an excessive blue luminosity with a greater specific star formation rate, which is in compliance with previous studies. This can be understood by investigating the mechanism which leads to lopsidedness in the disk. For galaxies in denser environments, tidal interactions or minor mergers are the two possible scenarios for generating lopsidedness in the disk, which leads to enhanced star formation. In the field environment where the interactions are less frequent, numerical studies have shown asymmetric gas accretion and a subsequent star formation can create strong lopsidedness in the disk. Hence, even though lopsidedness is a ubiquitous phenomenon observed in both groups or isolated environments, the formation mechanism and the lifetime for the $m=1$ mode in galaxies may depend on the environment within which the galaxies reside. We also verify the results obtained from the previous studies that lopsided galaxies have lower concentration index and are less massive compared to the symmetric galaxies. Finally, we discuss the dependence of the performance of the trained model on key parameters like redshift and different spiral morphologies.\\

\section*{Acknowledgments}
We thank the anonymous referee for their suggestions, which
helped to improve the clarity of the paper. \\

We thank Prime Minister’s Research Fellowship (PMRF ID - 0903060) for funding this project. We acknowledge discussions with Dr. Dmitry Makarov and Dr. Sergey Savchenk and we thank them very much for their valuable suggestions. We are also grateful to Prof. Françoise Combes for very useful suggestions during the initial phase of the work. We also thank Mr. Ganesh Narayanan for the useful comments and suggestions.\\

Funding for the Sloan Digital Sky Survey V has been provided by the Alfred P. Sloan Foundation, the Heising-Simons Foundation, the National Science Foundation, and the Participating Institutions. SDSS acknowledges support and resources from the Center for High-Performance Computing at the University of Utah. SDSS telescopes are located at Apache Point Observatory, funded by the Astrophysical Research Consortium and operated by New Mexico State University, and at Las Campanas Observatory, operated by the Carnegie Institution for Science. The SDSS web site is \url{www.sdss.org}.\\

SDSS is managed by the Astrophysical Research Consortium for the Participating Institutions of the SDSS Collaboration, including Caltech, The Carnegie Institution for Science, Chilean National Time Allocation Committee (CNTAC) ratified researchers, The Flatiron Institute, the Gotham Participation Group, Harvard University, Heidelberg University, The Johns Hopkins University, L’Ecole polytechnique f\'{e}d\'{e}rale de Lausanne (EPFL), Leibniz-Institut f\"{u}r Astrophysik Potsdam (AIP), Max-Planck-Institut f\"{u}r Astronomie (MPIA Heidelberg), Max-Planck-Institut f\"{u}r Extraterrestrische Physik (MPE), Nanjing University, National Astronomical Observatories of China (NAOC), New Mexico State University, The Ohio State University, Pennsylvania State University, Smithsonian Astrophysical Observatory, Space Telescope Science Institute (STScI), the Stellar Astrophysics Participation Group, Universidad Nacional Aut\'{o}noma de M\'{e}xico, University of Arizona, University of Colorado Boulder, University of Illinois at Urbana-Champaign, University of Toronto, University of Utah, University of Virginia, Yale University, and Yunnan University.\\

\textit{Software}: astropy (\citealp{astropy..2013,astropy..2018}), pytorch (\citealp{Pytorch..2019}), seaborn (\citealp{seaborn}).
\bibliographystyle{aa}
\bibliography{CNN_bibliography}

\end{document}